\pdfoutput=1
\documentclass[11pt]{article}
\usepackage[T1]{fontenc}
\usepackage{lmodern}
\usepackage[a4paper,margin=2.5cm]{geometry}
\usepackage{amsmath,amssymb,amsthm}
\usepackage{booktabs}
\usepackage{array}
\usepackage{graphicx}
\usepackage[numbers,sort&compress]{natbib}
\usepackage{xcolor}
\usepackage[colorlinks=true,linkcolor=blue,citecolor=blue,urlcolor=blue]{hyperref}
\usepackage{microtype}

\newtheorem{lemma}{Lemma}
\newcommand{\tv}{\operatorname{TV}}
\newcommand{\Dob}{\tau}
\newcommand{\lamtwo}{\lvert\lambda_2\rvert}
\newcommand{\Dros}{\textit{Drosophila}}

\title{Pre-registered spectral and certified mixing analysis of the male \Dros{} central nervous system connectome}
\author{Eran Kopel\\[2pt] \small Tel Aviv University, Tel Aviv, Israel\\ \small\texttt{erankopel@tauex.tau.ac.il}}
\date{September 2026}

\begin{document}
\maketitle

\begin{abstract}
We report a pre-registered test of five hypotheses about the synapse-flow random walk on the
connectome of the male \Dros{} central nervous system (166{,}700 neurons, 124 million synapses), in
which a walker moves to a postsynaptic partner in proportion to synapse number. The hypotheses
concern the spectral gap and its dependence on the neck connective, the localisation of the leading
modes of an input-normalised signed map, the certified mixing depth measured by the Dobrushin
coefficient, its robustness to per-synapse detection confidence, and the dependence of the signed
spectrum on the neurotransmitter sign convention. After public registration and a code freeze, each
was tested once. Two held: the two-neuron photoreceptor modes that lead the signed spectrum are an
artefact of normalising by very small inputs, removed by an input floor at the median, and they do
not depend on the sign convention. Three failed: at a threshold of five synapses per connection the
spectral gap is an order of magnitude smaller than predicted ($\lamtwo=0.9965$) and the chain is
certifiably far from mixed after 128 steps, and at no confidence level tested was the mixing depth
certifiably robust. Exploratory analyses trace the slow mode to eleven neurons of the cord, seven of
them mesothoracic efferent neurons, that become nearly closed once weak connections are dropped,
find a similar boundary set in the lamina of the FlyWire female brain, show that null models
preserving cell-type wiring reproduce the slow modes, and show that exact chains built from the
confidence data change the depth profile after 32 steps by at most 0.017 at the pre-registered
confidence level, far less than the worst-case certificate allows. Random-walk summaries of
connectomes are dominated by nearly closed sets at the edges of the reconstruction unless these are
treated explicitly.
\end{abstract}

\section{Introduction}\label{sec:intro}

Whole-nervous-system connectomes invite dynamical summaries. Random walks on the wiring diagram
identify neurons that attract or repel flow \cite{lin2024}; eigen-decompositions of linear models
built from synapse counts propose ``eigencircuits'' as candidate units of computation
\cite{pospisil2024}; spiking models built from synapse counts reproduce sensorimotor responses
\cite{shiu2024}; input-normalised influence models, maximum-flow analyses and layer
assignments describe how sensory signals reach motor neurons \cite{schlegel2021,winding2023,bates2026,berg2026}.
These summaries are usually reported as point estimates computed once, without a prediction stated
in advance, without a guarantee on their numerical accuracy, and without a test of how much they
depend on the uncertainty of the reconstruction itself, although automated synapse detection
assigns every synapse a confidence score \cite{buhmann2021,berg2026}.

The complete connectome of the male \Dros{} central nervous system \cite{berg2026,malecnsdata},
which joins the brain and the ventral nerve cord (VNC), makes a whole-CNS version of these questions
possible. (A female brain-and-cord connectome, BANC, appeared shortly before \cite{bates2026}; it was
analysed with an unsigned influence model, spectral clustering and centrality, not with the mixing
questions asked here.) We ask a simple question of the synapse-flow random walk, in which a walker
at a neuron moves to a postsynaptic partner with probability proportional to the number of
synapses: after how many synaptic steps has the walk forgotten where it started? The quantity that
answers it is the Dobrushin ergodicity coefficient $\Dob(P^r)$ of the $r$-step chain
\cite{dobrushin1956,seneta2006}, the largest total-variation distance that two starting neurons can
still keep after $r$ steps \cite{levin2017}. The number of steps after which it becomes small is the
classical counterpart of the entanglement-breaking index of a quantum channel, the number of
self-compositions after which a channel forgets its input \cite{kopel2026a}. Unlike a spectral gap, $\Dob(P^r)$ at finite $r$ can be bounded from both sides
by certified numerical bounds, and the bounds can be carried through a stated class of
reconstruction errors.

We did this as a pre-registered study \cite{nosek2018}. An exploratory analysis of the male CNS
(19 September 2026) produced five hypotheses with numerical pass and fail criteria, about the
spectral gap of the chain and its dependence on the neck connective (H1), the localisation of the
leading modes of an input-normalised signed linear map (H2), the certified mixing depth at a
threshold of five synapses per connection (H3), the robustness of the certified depth of the
one-synapse chain to per-synapse confidence (H4), and the dependence of the signed spectrum on the neurotransmitter sign convention
(H5). The pre-registration was made public on 26 September 2026 \cite{kopel2026reg}, the
confirmatory scripts were tested on synthetic data only and frozen \cite{kopel2026freeze}, and each
hypothesis was then run once. Section~\ref{sec:confirm} reports all five against their frozen
criteria: H2 and H5 held, H1, H3 and H4 failed. Section~\ref{sec:explore} reports analyses that we
made afterwards to understand the failures; they are exploratory and labelled as such. They show
that the slowest mode of the chain at the five-synapse threshold is carried by eleven neurons of the
cord, seven of them efferent neurons of the mesothoracic neuromere, that become a near-absorbing set
once weak connections are dropped, that the same kind of boundary set carries the slowest mode of the FlyWire female brain at one
synapse \cite{dorkenwald2024,schlegel2024}, and that chains built from the confidence data change the depth
profile far less than the worst-case certificate allows, most of whose slack lies in the step-by-step
accumulation of perturbations rather than in the radius assigned to each neuron.
\section{Data and methods}\label{sec:methods}

\subsection{Connectome and neuron set}

We used the male CNS connectome version~1.0, flat-connectome release at minimum synapse confidence
0.5 \cite{berg2026,malecnsdata} (CC-BY). The three input tables were checked against md5 sums frozen
in the pre-registration. Neurons are the 166{,}700 bodies with a superclass annotation; the
segment-to-segment table (151{,}856{,}684 rows, 311{,}833{,}243 synapses) restricted to pairs of
neurons gives 25{,}582{,}938 directed edges carrying 124{,}177{,}617 synapses, 6{,}242{,}118 of
them with at least five synapses. The giant strongly connected component (SCC) at weight $w\ge 1$,
denoted $G$, has 165{,}314 neurons (99.2\,\%) and 25{,}545{,}360 edges; at $w\ge5$ the giant SCC
has 157{,}821 neurons and 6{,}120{,}084 edges. Signs follow the consensus neurotransmitter
prediction of each neuron \cite{berg2026}: acetylcholine $+1$; GABA, glutamate and histamine $-1$;
all others $0$ (base convention). The alternative convention of Pospisil et al.\ \cite{pospisil2024}
adds dopamine $+1$, serotonin and octopamine $-1$.

\subsection{Synapse-flow chain and signed map}

On a giant SCC with weights $w_{ij}$ (synapses from $i$ to $j$), the synapse-flow chain is
$P_{ij}=w_{ij}/\sum_k w_{ik}$, a row-stochastic matrix; a walker at $i$ moves to a postsynaptic
partner in proportion to synapse number. Its stationary distribution $\pi$ was computed by power
iteration and its leading eigenvalues by ARPACK \cite{lehoucq1998} (implicitly restarted Arnoldi,
tolerance $10^{-8}$, seeded starting vectors). The signed map is $M_{ji}=s_i w_{ij}/\mathrm{in}_j$,
with $s_i$ the sign of the presynaptic neuron and $\mathrm{in}_j$ the input weight of $j$ from
partners of nonzero sign (postsynaptic normalisation); the floored map of H2 replaces
$\mathrm{in}_j$ by $\max(\mathrm{in}_j,F)$. The participation ratio of an eigenvector $v$ is
$1/\sum_i p_i^2$ with $p_i=|v_i|^2/\sum_k |v_k|^2$.

\subsection{Mixing depth and certified bounds}

For a stochastic matrix $Q$, the Dobrushin coefficient is
\begin{equation}
  \Dob(Q)=\tfrac12\max_{a,b}\lVert Q_{a\cdot}-Q_{b\cdot}\rVert_1 = 1-\min_{a,b}\sum_k \min(Q_{ak},Q_{bk}),
\end{equation}
so that $\Dob(P^r)=0$ exactly when every starting neuron leads to the same distribution after $r$
steps \cite{dobrushin1956,seneta2006}. Two certified bounds are cheap to compute on a chain of this
size. For any set $W$ of witness rows,
$\Dob(P^r)\ge\max_{a,b\in W}d_r(a,b)$ with $d_r(a,b)=\tfrac12\lVert(e_a-e_b)P^r\rVert_1$; and for any
set $K$ of columns, $\Dob(P^r)\le 1-\sum_{k\in K}\min_i (P^r)_{ik}$ \cite{seneta2006}. Witnesses were
90 random rows and the 10 rows of largest stationary mass (seed 7); hub columns were the 500 columns
of largest stationary mass. Rounding was bounded a priori \cite{higham2002}: after $r$ steps each
propagated row carries an $\ell_1$ error of at most about $r(D+1)u$, with $D$ the largest in- or
out-degree and $u=2^{-53}$, and the final sums add at most about $2nu$; a slack of $10^{-9}r$,
subtracted from lower and added to upper bounds, covers both (and the rounding of the robust sums
below) with a margin of at least a factor of ten at every $r$. The bounds are therefore statements about the exact chain defined by the data; in the text and
tables they are rounded to the digits shown, and the result files give them in full.

\subsection{Perturbation classes}

A perturbation class is a set of chains $P'$ that the data cannot rule out. The \emph{uniform}
class $C(\delta)$ multiplies every edge weight by a factor in $[1-\delta,1+\delta]$. The
\emph{confidence} class at level $c$ allows every edge weight $w'_{ij}$ between $w^{\mathrm{hi}}_{ij}(c)$,
the number of its synapses with $\min(\text{conf}_\text{pre},\text{conf}_\text{post})\ge c$, and
$w_{ij}$; that is, any subset of the synapses below confidence $c$ may be false detections. Both
classes are handled by the following two facts (proofs in Appendix~\ref{app:proofs}).

\begin{lemma}\label{lem:row}
If $0\le w'\le w$ entrywise on a row with total weights $W'=\sum_j w'_j>0$ and $W=\sum_j w_j$, the
normalised rows satisfy $\tv(w/W,\,w'/W')\le (W-W')/W$.
\end{lemma}

\begin{lemma}\label{lem:robust}
Let $P,P'$ be stochastic with $\tv(P_{i\cdot},P'_{i\cdot})\le\varepsilon_i$ for every row $i$, and
$x_t=(e_a-e_b)P^t$. Then
$\bigl|d_r(a,b;P')-d_r(a,b;P)\bigr|\le\sum_{t=0}^{r-1}\sum_i |x_t(i)|\,\varepsilon_i$.
\end{lemma}

For the confidence class, Lemma~\ref{lem:row} gives
$\varepsilon_i(c)=1-W^{\mathrm{hi}}_i(c)/W_i$, the fraction of the row's synapses below confidence
$c$, and Lemma~\ref{lem:robust} turns every witness pair into a lower bound on $\Dob(P'^r)$ valid for
the whole class. The exploratory analysis had used this argument without the $t=0$ term; that
error, found before registration, is corrected throughout (Section~\ref{sec:prereg}).

\subsection{Pre-registration and protocol}\label{sec:prereg}

The pre-registration was drafted and frozen on 19 September 2026 after an exploratory phase in which
every number of its Section~4 was computed; it was first made public, byte-identical, on
26 September 2026 \cite{kopel2026reg}, which is its registration date. Before any confirmatory
computation, a dated log of deviations and clarifications fixed every definition the text left open
(the removal sets of H1, the floor of H2, the graph, witnesses and confidence grid of H4, the
carrier sets of H5, the eigen-solver seeds and the run protocol), and the confirmatory scripts were
tested only on synthetic data against independent dense computations and frozen
\cite{kopel2026freeze}. The pre-registration text contained three errors, found and logged before
registration: the robust lower bounds printed in its Section~4.4 had been computed without the
$t=0$ term of Lemma~\ref{lem:robust} and were too high by $0.21$; its knock-out of ascending and
descending neurons removed 3{,}693 neurons, not the 3{,}160 stated; and its claim that no
brain-and-cord connectome had been analysed before overlooked BANC \cite{bates2026}. The analysis
ran once, on 26 September 2026, on a Linux container (two cores, 7.8\,GB) with Python 3.11,
NumPy 2.4.4, SciPy 1.17.1 and one BLAS thread, after a check that every executed file matched the
frozen manifest; every script ran once, without error.
\section{Confirmatory results}\label{sec:confirm}

Table~\ref{tab:verdicts} lists every pre-registered criterion with the value obtained and the
verdict. The reference values of the exploratory phase (at $w\ge1$) were reproduced to every digit
printed in the pre-registration before the confirmatory run.

\begin{table}[t]
\centering
\small
\caption{The five pre-registered hypotheses, each run once after the freeze. $\lamtwo$: modulus of
the second eigenvalue of the synapse-flow chain on the giant SCC; $\rho$: spectral radius of the
signed map; PR: participation ratio; $\Dob(P^r)$: Dobrushin coefficient of the $r$-step chain.}
\label{tab:verdicts}
\begin{tabular}{@{}l>{\raggedright\arraybackslash}p{7.5cm}>{\raggedright\arraybackslash}p{3.4cm}l@{}}
\toprule
 & Criterion (frozen) & Value & Met \\
\midrule
H1 (i) & $\lamtwo$ at $w\ge5$ lies in $[0.90,0.97]$ & $0.996524$ & no \\
H1 (ii) & removing ascending, descending and sensory-ascending neurons gives $1-\lamtwo<0.002$ & $0.000190$ & yes \\
H1 (iii) & removing the motor neurons moves $\lamtwo$ by less than $0.01$ & $0.000726$ & yes \\
\textbf{H1} & & & \textbf{fail} \\
\addlinespace
H2 (a) & with the floor $F$ (median input, 342 synapses), $\rho<0.85$ & $0.83484$ & yes \\
H2 (b) & at least 10 of the top 20 modes have PR $>20$ & 19 of 20 & yes \\
H2 (c) & descriptive: a top-20 mode with more than half its power on lobula-plate or ellipsoid-body neurons & mode 1, 99.3\,\% on the ellipsoid body & reported \\
\textbf{H2} & & & \textbf{pass} \\
\addlinespace
H3 (1) & certified lower bound on $\Dob(P^{32})$ at $w\ge5$ exceeds $0.20$ & $0.9229$ & yes \\
H3 (2) & first checkpoint with lower bound $<0.05$ lies in $[48,96]$ & none up to 128 ($0.6695$ at 128) & no \\
\textbf{H3} & & & \textbf{fail} \\
\addlinespace
H4 & certified lower bound on $\Dob(P'^{32})$ over the confidence class at $c=0.70$ exceeds $0.10$ & no certificate ($-0.52$) & no \\
H4 (clause) & largest level $c\in[0.51,0.95]$ at which the bound at $r=32$ exceeds $0.10$ & none (best: $0.071$ at $c=0.51$) & reported \\
\textbf{H4} & & & \textbf{fail} \\
\addlinespace
H5 & $\rho$ changes by less than $0.02$ under the Pospisil convention & $0.000015$ & yes \\
H5 & the carriers of the top six modes are unchanged & same six neurons & yes \\
\textbf{H5} & sensitivity result & & \textbf{pass} \\
\bottomrule
\end{tabular}
\end{table}

\paragraph{H1: the gap at five synapses is an order of magnitude smaller than predicted.}
At $w\ge1$ the chain has $\lamtwo=0.94503$, a spectral gap of $0.055$. At $w\ge5$ the prediction
was a gap between $0.03$ and $0.10$; the observed value is $\lamtwo=0.996524$, a gap of $0.0035$.
The two knock-out predictions held: without the ascending, descending and sensory-ascending neurons
the giant SCC (153{,}784 neurons) is still two-sided, held together by 18 residual neck neurons, and
has $1-\lamtwo=0.00019$; without the motor neurons $\lamtwo$ moves by $0.0007$.

\paragraph{H2 and H5: the photoreceptor modes are a normalisation effect.}
The leading eigenvalues of the signed map come in near-equal pairs of opposite sign
($-0.9151$, $+0.9137$, $-0.9122$, $+0.9098$), each mode carried by a pair of R7/R8 photoreceptors
(participation ratio 2.0 to 2.4). These photoreceptors receive few synapses (the six carriers have
2 to 43 input synapses from partners of nonzero sign, against 63 to 174 outputs), so postsynaptic
normalisation turns a reciprocal pair into a two-neuron loop of gain close to one. With the input floored at the
median, $F=342$, the spectral radius falls to $0.83484$ and 19 of the 20 leading modes extend over
more than 20 neurons. The leading floored mode is carried by GABAergic ellipsoid-body ring neurons
(ER3d\_b and ER3p\_a, participation ratio 96.6, 99.3\,\% of its power on neurons whose primary
neuropil is the ellipsoid body), and it is the ninth mode of the unfloored map with the same
eigenvalue: the floor removes the photoreceptor loops above it and leaves it untouched. Six of the
next eight modes are ring-neuron modes as well (ER4d, ER3p\_a, ER3m and three on ER3w\_b) and two
are on the leg sensory neurons LgAG1; the remaining modes of the top 20 are on Kenyon cells,
antennal-lobe projection neurons and sensory neurons of the cord. The result is the same with the floor at
700 synapses (18 of 20 modes extended). Under the Pospisil sign convention the unfloored spectral
radius changes by $1.5\times10^{-5}$ and the same six neurons carry the top six modes (H5).

\paragraph{H3: certified slow mixing.}
At $w\ge5$ the certified bounds are $0.923\le\Dob(P^{32})\le0.967$ and
$0.669\le\Dob(P^{128})\le0.816$ (Figure~\ref{fig:depth}): after 128 synaptic steps two starting
neurons can still be told apart with total-variation distance at least $0.669$. The first criterion held and the second failed,
because the bound never dropped below $0.05$. This is consistent with H1: at $\lamtwo=0.9965$ a
deviation decays by a factor ten only after several hundred steps.

\paragraph{H4: the worst-case certificate does not survive the confidence class.}
All 123{,}954{,}262 synapse-partner rows inside $G$ matched the connection weights exactly, and none
had a confidence below $0.5$. On average 14.4\,\% of a neuron's output synapses (12.3\,\% weighted by
the stationary distribution) have $\min(\text{conf}_\text{pre},\text{conf}_\text{post})<0.7$. With
per-row radii of this size, the certified lower bound of Lemma~\ref{lem:robust} is $0.821$ at $r=2$
and $0.396$ at $r=4$ and gives no certificate from $r=8$ on (Table~\ref{tab:h4}). Even at $c=0.51$,
where on average 0.5\,\% of a neuron's synapses lie below the level, the bound is $0.071$ at $r=32$, short of the
pre-registered $0.10$. As the pre-registration requires us to state: the depth result is not
certifiably robust to reconstruction uncertainty at the 0.7 level, nor at any level tested.

\begin{table}[t]
\centering
\small
\caption{H4: certified lower bounds on $\Dob(P'^r)$ over the confidence class at level $c$, and over
the uniform class $C(0.05)$, with 100 witness rows on $G$. A negative value means that no certificate
is obtained. The last row is the nominal bound for $P$ itself.}
\label{tab:h4}
\begin{tabular}{@{}lrrrrrrr@{}}
\toprule
$r$ & 1 & 2 & 4 & 8 & 16 & 32 & 48 \\
\midrule
$c=0.51$ & 1.000 & 0.997 & 0.977 & 0.860 & 0.555 & 0.071 & $-0.015$ \\
$c=0.55$ & 1.000 & 0.984 & 0.894 & 0.595 & $-0.028$ & $-0.071$ & $-0.071$ \\
$c=0.60$ & 1.000 & 0.941 & 0.759 & 0.264 & $-0.192$ & $-0.198$ & $-0.199$ \\
$c=0.70$ & 1.000 & 0.821 & 0.396 & $-0.470$ & $-0.511$ & $-0.520$ & $-0.522$ \\
$C(0.05)$ & 0.789 & 0.579 & 0.156 & $-0.441$ & $-0.481$ & $-0.502$ & $-0.508$ \\
\addlinespace
nominal $P$ & 1.000 & 1.000 & 0.998 & 0.936 & 0.707 & 0.312 & 0.127 \\
\bottomrule
\end{tabular}
\end{table}
\section{Exploratory analyses}\label{sec:explore}

Everything in this section was computed after the confirmatory run, to understand its outcome. None
of it was pre-registered, and none of it changes a verdict of Table~\ref{tab:verdicts}. The scripts,
logs and result files are in the repository under \texttt{code/exploratory} and
\texttt{exploratory/}.

\subsection{What carries the slow mode}\label{sec:slow}

For the chain on each giant SCC we computed the left and right eigenvectors of $\lambda_2$ and the
sweep cut of smallest conductance along the right eigenvector, a standard way of locating a
metastable set \cite{chung2005,deuflhard2005}: the set $S$ of neurons with the smallest (or
largest) eigenvector entries that minimises
$\phi(S)=F(S\to S^c)/\min\{\pi(S),\pi(S^c)\}$, with $F$ the stationary probability flow per step.

At $w\ge5$ the slow mode is carried by eleven neurons of the VNC (Table~\ref{tab:trap}; VNC type
names follow the systematic nomenclature of the male adult nerve cord connectome
\cite{takemura2024,marin2024,cheong2026}): seven efferent neurons with somata on the midline of the mesothoracic neuromere (EN00B001, EN00B008,
two EN00B011, two EN00B015 and the ventral unpaired median neuron mesVUM-MJ), two IN03B088 and one
IN19A061 interneuron, and one abdominal motor neuron, MNad21. Together they hold 12.8\,\% of the
stationary mass of the chain, a walker leaves them with probability $3.5\times10^{-3}$ per step,
and the two-block estimate $1-p-q=0.99598$ reproduces $\lamtwo=0.99652$. The left eigenvector
puts 99.7\,\% of its negative mass in the cord. The eleven neurons are not a trap at $w\ge1$: there
they hold 0.4\,\% of the stationary mass and a walker leaves them with probability 0.29 per step.
What changes is their output. At $w\ge1$ they send 299 synapses to the other neurons of $G$, only 43 of
them on connections of five or more synapses (six connections), while they receive 21{,}846 synapses,
19{,}378 of them on such connections, and 514 synapses connect them with each other. Dropping weak
connections removes most of the way out and little of the way in. Efferent neurons send their main output out of the CNS, to targets that are not in the
connectome, so the within-CNS normalisation of the chain leaves them with few, mostly local, exits.
This set also sets the certified bound of H3: from $r=4$ on, the witness pair that attains the
bound has one member inside it.

\begin{figure}[t]
\centering
\includegraphics[width=0.82\linewidth]{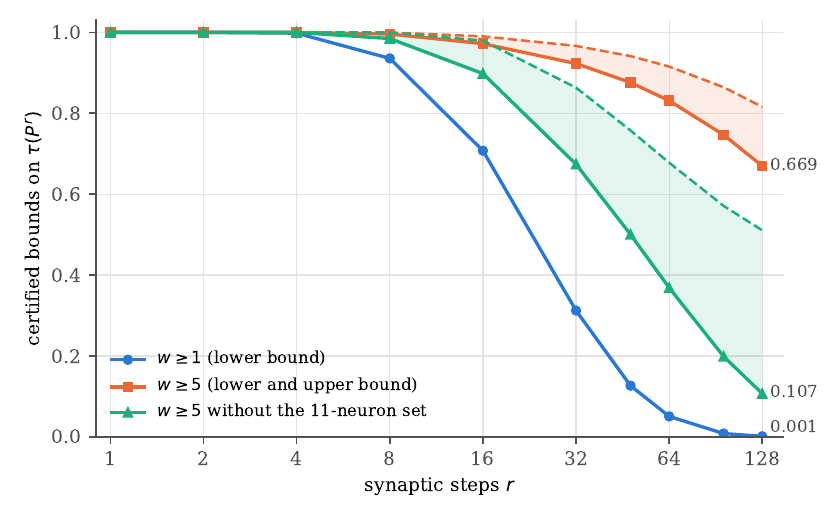}
\caption{Certified bounds on the Dobrushin coefficient $\Dob(P^r)$ of the synapse-flow chain of the
male CNS: the largest total-variation distance between two starting neurons after $r$ synaptic
steps. Blue: witness lower bound at $w\ge1$ (the nominal chain of H4); orange: lower (solid) and
upper (dashed) bounds at $w\ge5$ (H3); green: the same at $w\ge5$ with the eleven-neuron set of
Table~\ref{tab:trap} removed (exploratory, Section~\ref{sec:slow}). Every bound holds for the exact
chain, with rounding accounted for; the true value lies between the lower and upper curve.}
\label{fig:depth}
\end{figure}

Removing the eleven neurons lowers $\lamtwo$ at $w\ge5$ to 0.9806, and the next metastable set is
again small and in the cord (five VNC interneurons and one motor neuron, 4.0\,\% of the stationary
mass, conductance 0.027). The certified bounds become $0.674\le\Dob(P^{32})\le0.863$ and
$0.107\le\Dob(P^{128})\le0.511$. Removing every efferent superclass instead (24 neurons of the SCC)
gives nearly the same numbers ($\lamtwo=0.9806$, the same six-neuron set next).
At the five-synapse threshold, then, the slowest dynamics of the synapse-flow chain are set by a
succession of small, nearly closed sets in the cord, not by the neck connective
(Figure~\ref{fig:depth}).

At $w\ge1$ the picture of the exploratory phase holds for the left eigenvector (94.3\,\% of one sign
in the brain, 92.9\,\% of the other in the cord), but no small set dominates: the sweep cut of
smallest conductance is a set of 727 neurons of the central complex and the anterior visual pathway
(414 with primary neuropil in the ellipsoid body; medulla-tubercle, tubercle-bulb, ring, EPG, PEN
and PEG neurons \cite{hulse2021}), with conductance 0.116 and a two-block estimate of 0.883 that does
not reproduce $\lamtwo=0.945$.

\begin{table}[t]
\centering
\small
\caption{The slow mode of the synapse-flow chain. Cut: smallest-conductance sweep set along the slow
right eigenvector; mass: its stationary probability; escape: probability per step that a walker in
the set leaves it (stationary flow divided by mass).}
\label{tab:trap}
\begin{tabular}{@{}lrrrr>{\raggedright\arraybackslash}p{3.8cm}@{}}
\toprule
Chain & $\lamtwo$ & cut size & mass & escape & carried by \\
\midrule
male CNS, $w\ge1$ & 0.94503 & 727 & 0.0047 & 0.116 & central complex, anterior visual pathway \\
male CNS, $w\ge5$ & 0.99652 & 11 & 0.1276 & 0.0035 & mesothoracic efferent neurons \\
male CNS, $w\ge5$, without the 11 & 0.98062 & 6 & 0.0398 & 0.027 & VNC interneurons, one motor neuron \\
FlyWire brain, $w\ge1$ & 0.99999 & 7 & 0.0001 & $2.8\times10^{-5}$ & photoreceptors R1--6, L2 \\
FlyWire brain, $w\ge5$ & 0.95463 & 891 & 0.0177 & 0.069 & anterior visual pathway, central complex \\
male brain only, $w\ge1$ & 0.92559 & 1{,}416 & 0.0179 & 0.114 & anterior visual pathway, central complex \\
male brain only, $w\ge5$ & 0.93179 & 1{,}495 & 0.0150 & 0.105 & anterior visual pathway, central complex \\
\bottomrule
\end{tabular}
\end{table}

\subsection{Is the failure of H4 the certificate or the data?}\label{sec:members}

Lemma~\ref{lem:robust} bounds every chain of the confidence class at once, by a first-order
argument that lets each step take its own worst case. To see how far the depth numbers actually move,
we computed specific chains exactly. $M(c)$ removes every synapse with confidence below $c$, a vertex
of the class at level $c$ (a neuron whose outputs would all be removed keeps them, as the class
allows); $B(k)$ keeps each synapse with probability equal to the midpoint of its confidence bin, a
random reconstruction that is not a member of the class. For each chain we recomputed the distances
$d_r$ of all 4{,}950 pairs of the H4 witness rows and $\lamtwo$ at both thresholds
(Table~\ref{tab:members}).

\begin{table}[t]
\centering
\small
\caption{Chains built from the confidence data (exploratory). $M(c)$: every synapse with confidence
below $c$ removed; $B(k)$: every synapse kept with probability equal to the midpoint of its confidence bin
(five draws).
Synapses: total weight kept; row TV: mean total-variation change of a neuron's output distribution;
set: how many of the eleven neurons of Table~\ref{tab:trap} are in the giant SCC at $w\ge5$;
$\max d_{32}$: the witness lower bound on $\Dob(P'^{32})$ for that chain; $\Delta_{32}$: largest
change of $d_{32}$ over the 4{,}950 witness pairs, relative to $P$.}
\label{tab:members}
\begin{tabular}{@{}lrrrrrrr@{}}
\toprule
Chain & synapses ($10^6$) & row TV & $\lamtwo$, $w\ge1$ & $\lamtwo$, $w\ge5$ & set & $\max d_{32}$ & $\Delta_{32}$ \\
\midrule
$P$ & 123.95 & 0.000 & 0.9450 & 0.9965 & 11 & 0.312 &  \\
\addlinespace
$M(0.60)$ & 116.58 & 0.042 & 0.9441 & 0.9964 & 11 & 0.304 & 0.008 \\
$M(0.70)$ & 106.16 & 0.080 & 0.9432 & 0.9963 & 11 & 0.296 & 0.016 \\
$M(0.80)$ & 86.76 & 0.134 & 0.9413 & 0.9403 & 1 & 0.286 & 0.048 \\
$M(0.90)$ & 60.84 & 0.210 & 0.9374 & 0.9521 & 0 & 0.264 & 0.074 \\
\addlinespace
$B(1)$ & 105.37 & 0.078 & 0.9440 & 0.9763 & 2 & 0.304 & 0.015 \\
$B(2)$ & 105.36 & 0.078 & 0.9439 & 0.9944 & 8 & 0.303 & 0.014 \\
$B(3)$ & 105.36 & 0.078 & 0.9439 & 0.9738 & 3 & 0.305 & 0.015 \\
$B(4)$ & 105.36 & 0.078 & 0.9438 & 0.9962 & 11 & 0.303 & 0.012 \\
$B(5)$ & 105.36 & 0.078 & 0.9441 & 0.9669 & 6 & 0.306 & 0.010 \\
\bottomrule
\end{tabular}
\end{table}

\begin{figure}[t]
\centering
\includegraphics[width=0.82\linewidth]{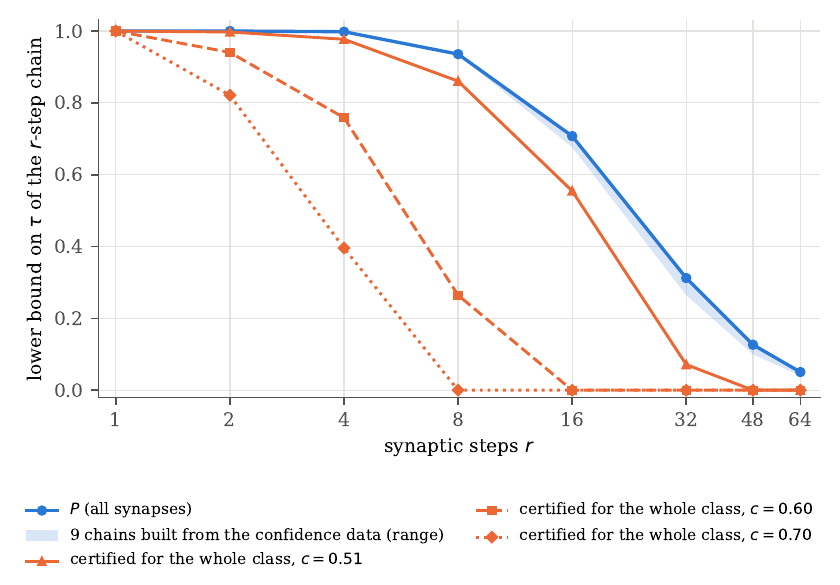}
\caption{The certificate of H4 against exact chains (exploratory). Blue line: witness lower bound
$\max_{a,b}d_r(a,b)$ on $\Dob(P^r)$ for the chain $P$ of all synapses ($w\ge1$, the 4{,}950 pairs of
the H4 witness rows); blue band: its range over the nine chains $M(0.60)$ to $M(0.90)$ and $B(1)$ to
$B(5)$ of Table~\ref{tab:members}. Orange: the certified lower bound of H4
(Lemma~\ref{lem:robust}), valid for every chain of the confidence class at level $c$, drawn at zero
where it is negative, that is, where no certificate is obtained. The chains $M(0.60)$ and $M(0.70)$
belong to the class at $c=0.70$.}
\label{fig:members}
\end{figure}

The chains move little (Table~\ref{tab:members}, Figure~\ref{fig:members}). Removing every synapse
below confidence 0.70 changes a neuron's output distribution by 0.080 in total variation on average;
it changes no witness distance at $r=32$ by more than 0.017 (and none at any $r\le64$ by more than
0.043), and $\max d_{32}$ falls from 0.312 to 0.296. Removing every synapse below 0.90, half of all
synapses, lowers $\max d_{32}$ to 0.264. The five random reconstructions remove 15\,\% of the synapses
and change no witness distance at $r=32$ by more than 0.016 (0.11 at $r=8$), while $\max d_r$ moves
by less than 0.01 at every $r$. $M(0.60)$ and $M(0.70)$ are members of the class at $c=0.70$, for
which the certified bound of H4 is negative from $r=8$ on. The second eigenvalue at $w\ge1$ stays
within 0.008 of its value in every chain. At $w\ge5$ it follows the eleven-neuron set of
Section~\ref{sec:slow}. In $M(0.80)$ and $M(0.90)$ the set keeps 15 and none of the 43 synapses on
its connections of five or more synapses to the other neurons of $G$, leaves the giant SCC (one of its
neurons remains in $M(0.80)$), and $\lamtwo$ falls to 0.940 and 0.952. The random reconstructions
keep 21 to 35 of those synapses; where the set stays in the giant SCC with most of its stationary
mass ($B(2)$ and $B(4)$, 8\,\% and 11\,\%) $\lamtwo$ is 0.994 and 0.996, and where it is broken up
(two to six of its neurons left, with mass at most 0.0014) $\lamtwo$ is 0.967 to 0.976, still
above its value at $w\ge1$. The slowest mode at five synapses is thus a property of a few
connections that detection errors can make or break.

Where the certificate loses can be located. Lemma~\ref{lem:row} charges row $i$ the fraction
$\varepsilon_i$ of its synapses below the level, which is the largest possible change of the row only
if all of them go to partners that receive no high-confidence synapse from $i$. The exact largest
change over the class has a closed form (proof in Appendix~\ref{app:proofs}).

\begin{lemma}\label{lem:sharp}
For a row with partner weights $w_j$, high-confidence parts $w^{\mathrm{hi}}_j$,
$l_j=w_j-w^{\mathrm{hi}}_j$ and $W=\sum_j w_j$, the largest total-variation change of the normalised
row over the confidence class is
\begin{equation*}
\varepsilon^*=\max_{S}\Bigl[\frac{w(S)}{W}-\frac{w^{\mathrm{hi}}(S)}{W-l(S)}\Bigr]\le\varepsilon,
\end{equation*}
the maximum over nonempty sets $S$ of partners (sums over $S$), attained by removing every
low-confidence synapse to $S$ and none to the other partners. (A row without high-confidence
synapses may be emptied, and then $\varepsilon^*=1$.)
\end{lemma}

We computed $\varepsilon^*$ exactly for rows with at most 12 partners and bounded it from above, still
rigorously, for the others. At $c=0.70$ the mean radius falls from 0.144 to 0.089, close to the 0.080
by which the vertex $M(0.70)$ actually moves a row, but the certificate gains little depth
(Table~\ref{tab:cert}): at $r=4$ the bound rises from 0.396 to 0.621, at $r=8$ it remains negative.
At $c=0.55$ it now certifies $\Dob(P'^{16})\ge0.106$ for the whole class, where the radius of H4 gave
nothing. At $r=32$ the best certificate, at $c=0.51$, rises from 0.071 to 0.096, still short of the
pre-registered 0.10, and from $c=0.52$ on there is none. Most of the slack is in
Lemma~\ref{lem:robust}. Evaluated with the actual row changes
of $M(0.70)$, which is not a certificate but the most favourable radius for that one chain, the lemma
gives $0.026$ at $r=8$, where $M(0.70)$ itself has $0.934$. The lemma adds, at every step, the
perturbations of all the rows that the two walkers visit, as if every one of them pushed the walkers
apart; in the actual chain they point in different directions and largely cancel.

\begin{table}[t]
\centering
\small
\caption{Where the certificate of H4 loses (exploratory). Certified lower bounds on $\Dob(P'^r)$ for
every chain of the confidence class at level $c$, with the row radius of H4 and with the radius of
Lemma~\ref{lem:sharp} (exact for rows with at most 12 partners, a certified upper bound
otherwise); for comparison, Lemma~\ref{lem:robust} evaluated with the actual row changes
of the vertex $M(c)$ (valid for that chain only), and the exact witness bound of $M(c)$ from
Table~\ref{tab:members}. Mean: mean radius over the rows of $G$. A negative value means that no
bound is obtained.}
\label{tab:cert}
\begin{tabular}{@{}llrrrrrr@{}}
\toprule
 & radius & mean & $r=2$ & $r=4$ & $r=8$ & $r=16$ & $r=32$ \\
\midrule
$c=0.55$ & $\varepsilon$ (Lemma~\ref{lem:row}, H4) & 0.028 & 0.984 & 0.894 & 0.595 & $-0.028$ & $-0.071$ \\
 & $\varepsilon^*$ (Lemma~\ref{lem:sharp}) & 0.023 & 0.987 & 0.916 & 0.665 & 0.106 & $-0.056$ \\
 & row change of $M(c)$, not certified & 0.023 & 0.987 & 0.918 & 0.671 & 0.119 & $-0.055$ \\
\addlinespace
$c=0.60$ & $\varepsilon$ (Lemma~\ref{lem:row}, H4) & 0.059 & 0.941 & 0.759 & 0.264 & $-0.192$ & $-0.198$ \\
 & $\varepsilon^*$ (Lemma~\ref{lem:sharp}) & 0.043 & 0.966 & 0.832 & 0.448 & $-0.140$ & $-0.152$ \\
 & row change of $M(c)$, not certified & 0.042 & 0.970 & 0.840 & 0.469 & $-0.132$ & $-0.143$ \\
& exact witness bound of $M(c)$ & & 1.000 & 0.998 & 0.935 & 0.702 & 0.304 \\
\addlinespace
$c=0.70$ & $\varepsilon$ (Lemma~\ref{lem:row}, H4) & 0.144 & 0.821 & 0.396 & $-0.470$ & $-0.511$ & $-0.520$ \\
 & $\varepsilon^*$ (Lemma~\ref{lem:sharp}) & 0.089 & 0.913 & 0.621 & $-0.065$ & $-0.304$ & $-0.321$ \\
 & row change of $M(c)$, not certified & 0.080 & 0.930 & 0.657 & 0.026 & $-0.265$ & $-0.281$ \\
& exact witness bound of $M(c)$ & & 1.000 & 0.998 & 0.934 & 0.697 & 0.296 \\
\addlinespace
nominal $P$ & & & 1.000 & 0.998 & 0.936 & 0.707 & 0.312 \\
\bottomrule
\end{tabular}
\end{table}

\subsection{Null models}\label{sec:null}

Each null model keeps every connection's presynaptic neuron and weight and re-draws its postsynaptic
partner by permuting targets within a group of connections, so out-degrees, out-strengths and the
multiset of in-connections are kept. N1 permutes over all connections; N2 within each pair
(superclass of source, superclass of target); N3 within each pair of cell types (three draws each,
Table~\ref{tab:null}). At $w\ge1$ the configuration null N1 mixes fast ($\lamtwo\approx0.15$, 9\,\%
of the stationary mass on the top 1\,\% of neurons against 42\,\% observed); preserving
superclass-to-superclass counts restores most of the slow mode ($\lamtwo=0.928$), and preserving
type-to-type counts restores all of it ($0.9446$ to $0.9447$ against $0.9450$, with a smallest-conductance cut of
714 to 716 neurons against 727 observed, and 41\,\% of the stationary mass on the top 1\,\%). At $w\ge5$ the superclass
null gives 0.91 to 0.94; the type null reproduces a trap of 11 to 13 neurons in two of three draws
($\lamtwo=0.998$ and $0.996$) and not in the third (0.953). The slow modes are thus a property of
cell-type wiring, and at $w\ge5$ they depend on the particular connections of a few efferent types.

\begin{table}[t]
\centering
\small
\caption{Null models: $\lamtwo$ of the giant SCC (three draws each) and stationary mass of the top
1\,\% of neurons.}
\label{tab:null}
\begin{tabular}{@{}lllll@{}}
\toprule
 & observed & N1 (configuration) & N2 (superclass blocks) & N3 (type blocks) \\
\midrule
$w\ge1$, $\lamtwo$ & 0.945 & 0.155, 0.152, 0.152 & 0.928, 0.928, 0.928 & 0.945, 0.945, 0.945 \\
$w\ge1$, top 1\,\% mass & 0.42 & 0.09 & 0.19 & 0.41 \\
$w\ge5$, $\lamtwo$ & 0.997 & 0.230, 0.224, 0.225 & 0.935, 0.913, 0.913 & 0.998, 0.996, 0.953 \\
$w\ge5$, top 1\,\% mass & 0.73 & 0.13 & 0.27 & 0.72, 0.69, 0.63 \\
\bottomrule
\end{tabular}
\end{table}

\subsection{The FlyWire female brain}\label{sec:flywire}

The same chain on the FlyWire whole-brain connectome of a female fly \cite{dorkenwald2024,schlegel2024}
(version 783, 139{,}255 proofread neurons, 54.5 million synapses \cite{flywire783}) and on the male
CNS with the cord removed (the brain-only row of H1) gives the numbers of Tables~\ref{tab:trap} and
\ref{tab:brains}. Two things agree between the datasets and one does not.

First, the slowest well-populated subsystem is the same in both brains: neurons of the anterior
visual pathway and the central complex \cite{hulse2021}. In FlyWire at $w\ge5$ the cut (891 neurons)
is led by medulla-tubercle (MeTu), tubercle-bulb (TuBu), ring (ER) and EPG neurons and Dm-DRA1; in
the male brain (1{,}416 and 1{,}495 neurons) by MeTu neurons, dorsal-rim R7 photoreceptors, Dm-DRA1
and EPG neurons, with Delta7 neurons at one synapse and Mi15 medulla neurons at five. The conductance of these sets is 0.07 to 0.11. The pathway from the dorsal
rim to the ellipsoid body carries sky-compass information; in the synapse-flow chain it is a weakly
coupled channel. Second, the certified depth profiles are of the same order: the lower bound
on $\Dob(P^{32})$ is 0.28 (FlyWire, $w\ge1$), 0.26 (FlyWire, $w\ge5$), 0.11 and 0.17 (male brain),
against 0.31 for the whole male CNS at $w\ge1$.

What does not agree is $\lamtwo$. In FlyWire at $w\ge1$ it is $0.99999$, set by seven neurons (four
R1--6 photoreceptors, one L2 and two unlabelled optic-lobe neurons) that hold about $10^{-4}$ of the
stationary mass and are left with probability $2.8\times10^{-5}$ per step. Like the efferent
neurons of the male cord, the set sits at an edge of the reconstruction, here the lamina. It
dominates $\lamtwo$ but, carrying almost no stationary mass, has no visible effect on the certified
depth profile. The second eigenvalue of a connectome random walk is a
statement about its most nearly closed set, however small; the depth profile is a statement about
where the walk actually goes.

\begin{table}[t]
\centering
\small
\caption{Brain-only chains: certified lower bounds on $\Dob(P^r)$ (100 witness rows, seed 7) and
the stationary mass of the top 1\,\% of neurons. The whole male CNS at $w\ge1$ is shown for
comparison.}
\label{tab:brains}
\begin{tabular}{@{}lrrrrrr@{}}
\toprule
Chain & neurons & $r=8$ & $r=16$ & $r=32$ & $r=64$ & top 1\,\% mass \\
\midrule
FlyWire brain, $w\ge1$ & 135{,}403 & 0.897 & 0.640 & 0.285 & 0.043 & 0.41 \\
FlyWire brain, $w\ge5$ & 119{,}756 & 0.933 & 0.631 & 0.261 & 0.050 & 0.56 \\
male brain only, $w\ge1$ & 145{,}035 & 0.854 & 0.451 & 0.112 & 0.007 & 0.35 \\
male brain only, $w\ge5$ & 139{,}150 & 0.895 & 0.549 & 0.165 & 0.016 & 0.48 \\
\addlinespace
male CNS, $w\ge1$ & 165{,}314 & 0.936 & 0.707 & 0.312 & 0.051 & 0.42 \\
\bottomrule
\end{tabular}
\end{table}
\section{Discussion}\label{sec:discussion}

\paragraph{What the pre-registered tests established.}
Two predictions held. The two-neuron photoreceptor modes that lead the spectrum of the
input-normalised signed map are an artefact of dividing by very small inputs (H2), and they do not
depend on the sign convention (H5). A floor at the median input removes them and leaves extended
modes of GABAergic ellipsoid-body ring neurons in the lead, the kind of circuit that Pospisil et al.\
found among the eigencircuits of the unnormalised female-brain matrix \cite{pospisil2024}. Linear
models and graph traversals that normalise by postsynaptic input \cite{schlegel2021,bates2026}
inherit this sensitivity at
their low-input neurons, which are often sensory; reporting the input weights of the carriers of any
leading mode is a cheap guard. Three predictions failed. The spectral-gap and depth predictions at
the five-synapse threshold (H1 (i), H3) were set by values of the weak-threshold chain, and the chain
at five synapses turned out to be an order of magnitude slower; the two knock-out predictions of
H1 (ii) and (iii) held. The certificate of robustness to reconstruction uncertainty
(H4) failed at every confidence level tested.

\paragraph{Boundary sets dominate slow modes.}
The exploratory analyses give one explanation for both depth failures, and it is a caution for any
random-walk analysis of a connectome. A random walk normalised within the reconstructed graph
treats every neuron as if all its outputs were in the graph. Efferent and motor neurons, whose main
targets are outside the CNS, and neurons at an incompletely reconstructed edge of the volume, are
left with a few within-graph exits; once weak connections are dropped these can close almost
completely. Eleven neurons of the male cord, seven of them mesothoracic efferent neurons, do this
at five synapses, and seven neurons at the lamina of the FlyWire brain do it at one. Such sets set $\lamtwo$ and any mixing time derived
from it, whatever their size, and they set certified depth bounds when they carry stationary mass.
Three remedies follow: add an explicit exit state that receives the synapses a neuron makes outside
the graph (the male CNS tables record synapses to non-neuron bodies, which makes this possible),
report results with boundary classes removed, and report depth profiles with the witness sets and
stationary masses that attain them rather than a single eigenvalue. The threshold matters as much:
the same connectome gives $\lamtwo=0.945$ at one synapse and $0.9965$ at five.

\paragraph{Certificates and reconstruction uncertainty.}
The failure of H4 is not evidence that the depth profile is fragile. Every chain that we built
from the confidence data, including vertices of the class that remove every synapse below the level,
moved the witness distances after 32 steps by at most 0.017 at the pre-registered level, and by less
than 0.08 even after half of all synapses were removed, where the certificate allows changes of
order one. Sharpening the row radius (Lemma~\ref{lem:sharp}) recovers only part
of the gap; most of it is lost in the step-by-step accumulation of Lemma~\ref{lem:robust}, which
cannot use the cancellation between the perturbations of different rows. The certificate cannot rule
out that some adversarial member of the class moves the profile much further, and whether one exists
is open. A useful certificate at the depths that matter would have to exploit that cancellation, or
restrict the class to what detection errors can plausibly do; until then, robustness of connectome
summaries to reconstruction uncertainty can be shown by resampling, as in Table~\ref{tab:members},
but not certified.

\paragraph{Limitations.}
The confidence class models false detections only; missing synapses, which incomplete proofreading
makes common \cite{berg2026}, are not in it, and a class that allowed them would be larger. Only the
bounds on $\Dob(P^r)$ are certified; eigenvalues were computed by ARPACK to a tolerance of
$10^{-8}$ and are not. The chains are not reversible, so the sweep cuts locate metastable sets
heuristically \cite{chung2005,deuflhard2005}. The FlyWire comparison differs in sex, reconstruction
pipeline and synapse detection. Most of all, the synapse-flow chain is a structural object: it says
how a walker that follows synapses spreads over the wiring diagram, not how activity spreads in a
fly.
\section*{Data and code availability}

The male CNS connectome v1.0 is public under CC-BY \cite{berg2026,malecnsdata}; the md5 sums of the
four input tables used here are recorded in the repository. The FlyWire connectivity data are on
Zenodo \cite{flywire783} and the FlyWire annotations in the flywire\_annotations repository
\cite{schlegel2024}. The pre-registration, the deviation log, the exploratory and confirmatory code,
the tests, the logs of the confirmatory run, all result files and the exploratory analyses of
Section~\ref{sec:explore} are at \url{https://github.com/erankopel/maleCNS-depth}; the registration
record and the frozen confirmatory analysis are archived at doi:10.5281/zenodo.22980248
\cite{kopel2026reg} and doi:10.5281/zenodo.22981189 \cite{kopel2026freeze}.

\section*{Acknowledgements}

This work used data made public by the FlyEM project team at HHMI Janelia and its collaborators
(male CNS) and by the FlyWire consortium; the author thanks them for releasing the data openly.
No external funding was received for this work. The author declares no competing interests.

\paragraph{Use of generative AI.}
This study and manuscript were prepared with the help of Claude (Anthropic).
It was used to write and run the exploratory, confirmatory and test scripts, to review the
confirmatory scripts before the freeze, to prepare the pre-registration record, the deviation log
and the reproduction and results notes, to check the references against Crossref, DataCite and the
publishers' pages, and to draft and revise the manuscript text. The author designed the study,
approved every pre-registration decision, directed and checked the work, verified the results, and
takes full responsibility for the content.

\bibliographystyle{unsrtnat}
\bibliography{references}

\appendix

\section{Proofs}\label{app:proofs}

\paragraph{Lemma~\ref{lem:row}.}
Write $p=w/W$ and $p'=w'/W'$. Since $W'\le W$, $p'_j=w'_j/W'\ge w'_j/W$, hence
$p_j-p'_j\le (w_j-w'_j)/W$ for every $j$. The total-variation distance is the sum of the positive
parts of $p_j-p'_j$, so $\tv(p,p')\le\sum_j (w_j-w'_j)/W=(W-W')/W$. If $W'=0$ the row of $P'$ may be
any distribution and $\tv\le1=(W-W')/W$ holds trivially. With $W'\ge W^{\mathrm{hi}}(c)$ for every
member of the confidence class, $\varepsilon_i(c)=1-W^{\mathrm{hi}}_i(c)/W_i$ bounds the row change
of every member. For the uniform class $C(\delta)$ every ratio $P'_{ij}/P_{ij}$ lies in
$[(1-\delta)/(1+\delta),(1+\delta)/(1-\delta)]$, which gives $\tv\le 2\delta/(1+\delta)$; the
pre-registered, slightly larger constant $\varepsilon=2\delta/(1-\delta)$ was kept.

\paragraph{Lemma~\ref{lem:robust}.}
Let $x'_t=(e_a-e_b)P'^t$ and $\Delta_t=x'_t-x_t$, so $\Delta_0=0$ and
$\Delta_{t+1}=\Delta_tP'+x_t(P'-P)$. Because $P'$ is stochastic, $\lVert yP'\rVert_1\le\lVert
y\rVert_1$ for every row vector $y$, and therefore
$\lVert\Delta_r\rVert_1\le\sum_{t=0}^{r-1}\lVert x_t(P'-P)\rVert_1$. Each term satisfies
$\lVert x_t(P'-P)\rVert_1\le\sum_i|x_t(i)|\,\lVert P'_{i\cdot}-P_{i\cdot}\rVert_1\le\sum_i|x_t(i)|\,2\varepsilon_i$.
Since $d_r=\tfrac12\lVert x_r\rVert_1$, the triangle inequality gives
$|d_r(P')-d_r(P)|\le\tfrac12\lVert\Delta_r\rVert_1\le\sum_{t=0}^{r-1}\sum_i|x_t(i)|\,\varepsilon_i$.
The $t=0$ term, $\varepsilon_a+\varepsilon_b$, is the one the exploratory script had omitted; the
test suite of the repository shows that without it the bound is violated by chains sampled from both
classes.

\paragraph{Lemma~\ref{lem:sharp}.}
For every member $w'$ of the class, $\tv(p,p')=\max_S\,(p(S)-p'(S))$ over sets $S$ of partners. For a
fixed $S$, $p'(S)=w'(S)/(w'(S)+w'(S^c))$ increases with $w'(S)$ and decreases with $w'(S^c)$, so its
smallest value over the class is $w^{\mathrm{hi}}(S)/(w^{\mathrm{hi}}(S)+w(S^c))
=w^{\mathrm{hi}}(S)/(W-l(S))$, attained by removing every low-confidence synapse to $S$ and none to
$S^c$. Taking the maximum over $S$ gives $\varepsilon^*$; and since $W-l(S)\le W$,
$w(S)/W-w^{\mathrm{hi}}(S)/(W-l(S))\le l(S)/W\le\varepsilon$. The maximum over the $2^d-1$ nonempty
sets was computed exactly for rows with $d\le12$ partners (3.6\,\% of the rows of $G$). For the
others we used an upper bound: for a fixed removed weight $C=l(S)$ the maximisation is a knapsack
problem in which a partner $j$ contributes $w_j/W-w^{\mathrm{hi}}_j/(W-C)$ at cost $l_j$; the ratio
of the two decreases with the high-confidence fraction $q_j=w^{\mathrm{hi}}_j/w_j$ for every $C$, so
the fractional relaxation is solved, for all $C$ at once, by taking partners in increasing order of
$q_j$, and its largest value along this path bounds $\varepsilon^*$ from above. On random test rows
the bound exceeded the exact value by at most $0.0053$ for 13 to 15 partners.

\end{document}